# Reply to Marchildon: absorption and non-unitarity remain well-defined in the Relativistic Transactional Interpretation

R. E. Kastner[*]
26 December 2017

ABSTRACT. I rebut some erroneous statements and attempt to clear up some misunderstandings in a recent set of critical remarks by Marchildon regarding the Relativistic Transactional Interpretation (RTI), showing that his negative conclusions regarding the transactional model are ill-founded.

In a recent paper, Marchildon (2018) makes some critical remarks regarding the ability of the ability of the Relativistic Transactional Interpretation to define absorption, and the accompanying non-unitary measurement transition, that are the hallmarks of the transactional picture. The purpose of this Letter is to rebut some erroneous statements and to clear up some misunderstandings therein, thereby demonstrating that Marchildon's negative conclusions regarding the status of RTI are unfounded.

First, Marchildon is apparently intent on finding empirical predictions from RTI that differ from standard quantum theory. However, as has been repeatedly stated in publications (and in private correspondence), RTI is empirically equivalent to standard QED (up to the non-unitary transition). *This is a theorem,* as noted by Davies (1972), discussed in Kastner and Cramer (2017.) *Empirically equivalent theories make identical empirical predictions.* Yet Marchildon continues to insist that in this case, they should not. The present author remains perplexed by this.

The only sense in which RTI differs from standard QM/QED is in predicting collapse (i.e. predicting that we will get definite outcomes, which we do in fact get). In contrast, the unitary-only theory fails to predict what we see; i.e., definite outcomes. Thus, to the extent that it differs from standard QM/QED (i,e only in predicting the measurement transition), RTI is empirically corroborated; while the unitary-only theory is not.

---

[*] rkastner@umd.edu; University of Maryland, College Park; Foundations of Physics Group

[1] Albeit at the relativistic level, involving particle creation and destruction.

[2] We disregard 'hidden variables' approaches here, since we are working with the bare theory (in the transactional picture).

[3] However the two factors arise from very different processes. In the virtual photon, line shift case, there are two interaction vertices corresponding to the transition from the base state to an

Prof. Marchildon next disputes the interpretation of the coupling constant *e* as the amplitude for emission or absorption of a real photon, asserting that "the charge is not associated with the amplitude of a physical process." But that is simply wrong. It is exactly contradicted by Feynman, the primary founder of QED, who said: "There is a most profound and beautiful question associated with the observed coupling constant, *e* – the amplitude for a real electron to emit or absorb a real photon…" (Feynman 1985, p. 129). The fact that each Feynman diagram represents a term in a sum in no way refutes this interpretation of the coupling amplitude. Such sums express situations in which no real photon was in fact emitted (usually because the photons are off-shell and/or their emission would violate the conservation laws). But the amplitude still functions as Feynman stated.

In fact, this issue has already been addressed in detail in the published literature, specifically in Kastner (2012a) and Kastner (2012b). It is shown in the former, Section 6.3.4, that the coupling amplitude plays an analogous role[1] with the amplitudes for 'which slit' detection in a two-slit experiment. Each 'which slit' component of the prepared quantum state |Ψ>, characterized by amplitude <A| Ψ > and <B| Ψ> respectively, corresponds to the *possibility* of passage through slit A or B, even though (with both slits open) such definite, detectable passage through a particular slit 'never happens as a real physical process.'[2] Of course, the amplitudes are added in this case (just as amplitudes for individual Feynman diagrams, involving products of amplitudes according to the Feynman rules, are added in the perturbation expansion); and they contribute an important quantitative component to the calculation for probability of final detection. The fact that the photon did not 'really go through slit A' (or slit B) does not in any way negate the fact that the quantity <A|Ψ> is still the amplitude for going through slit A! Yet denying that is what Marchildon's objection amounts to. If he wants to disallow the interpretation of the coupling amplitude as the amplitude for photon emission (or absorption), because such emissions or absorptions do not always occur, he must do the same for the 'which slit' amplitudes in a two-slit experiment. But of course nobody wants to do that, nor should they.

Moreover, the amplitude for emission/absorption of a photon does indeed correspond to a real physical process when photons are on-shell and conservation laws are satisfied. In this case, offer wave (OW) and confirmation wave (CW) generation take place; so the coupling amplitude is squared, along with the relevant transition probability for the decay process involved. So, just as in a 'which slit' detection situation, the emission/absorption of a real photon can indeed occur as a real physical process, under the right physical conditions and with the appropriate probabilities; cf. equations (10) and (11) in Kastner and Cramer (2017).

---

[1] Albeit at the relativistic level, involving particle creation and destruction.

[2] We disregard 'hidden variables' approaches here, since we are working with the bare theory (in the transactional picture).

In case the reader is still not convinced that coupling amplitudes are indeed amplitudes for emission or absorption of photons regardless of whether these processes 'really' happen, one may consult Sakurai's illuminating discussion of line shifts and line broadening (Sakurai 1973, pp. 66-7). Atomic energy level shifts (such as the Lamb shift) arise from virtual photon interactions, while the broadening of levels corresponds to the probability of emission (or absorption) of real photons. Both processes emerge naturally from a single integral expression for the amplitude of the intermediate (perturbed) atomic state, summed over all possible such states (cf. Sakurai 1973, eq (2.230)). In order to evaluate the integral containing the shift ΔE, a small imaginary part is added, which upon integration leads to two distinct expressions. One applies to off-shell, virtual photons only, and the other applies to on-shell, real photons that satisfy the conservation laws and are therefore emitted or absorbed with the calculated time-dependent probability (decay rate). Both such expressions are multiplied by the square of the relevant transition amplitude, which contains the coupling constant as a factor.[3] The coupling constant does not change its physical significance based on whether specific photons are off-shell or not; instead, it features in different calculations for each case. This is analogous to the fact that the amplitude for going through slit A, <A|Ψ>, does not change its physical significance based on whether or not both slits are open, but features in a different calculation for each case. Thus, it is clear that it is fully justified to interpret the coupling amplitude, describing charge, as the amplitude for emission or absorption of a real photon. This does not in any way require that it always be associated with such emission or absorption, any more than an amplitude to go through a slit must always be associated with detectably going through that slit.

Prof. Marchildon's next move is to question whether experiments with Buckeyballs are consistent with RTI (though again, they have to be, since RTI is empirically equivalent to the standard theory). The order-of-magnitude figures for the probabilities of non-unitary behavior in Kastner (2018) are intended to be only very rough estimates, since they neglect the relevant transition probabilities. Obtaining specific probabilities for the onset of non-unitarity for the mesoscopic realm (including Buckeyballs) requires detailed calculations based on the specific structure of whatever molecules are being used, and those calculations will be done with standard QM (*with which RTI is empirically equivalent*). A molecule that for example is subject to excitation by extraneous photons will be a source of loss of unitarity (leading to 'which-way information') even according to standard QM. It's just that standard QM won't be able to explain why (at least not in terms of a non-unitary measurement transition at that point). So once again, Marchildon's apparent hope that RTI will be refuted by experiments is misguided, since it is empirically equivalent to standard theory. Of course, anyone who wants to check this can certainly do so, using the correct detailed calculations for the systems under study; i.e. using not just the square of the coupling amplitude alone, as Marchildon seems to be assuming, but the total probability (decay rate) which includes the relevant transitions between states of the molecules.

---

[3] However the two factors arise from very different processes. In the virtual photon, line shift case, there are two interaction vertices corresponding to the transition from the base state to an intermediate state and back again (summed over all possible intermediate states). In the real photon case, the squaring arises from the confirming (adjoint) response of each absorber to the emitter (summed over all possible final states). The distinction between these two cases is discussed in Sakurai (1973), p. 67 (although not within the transactional model).

Marchildon's next objection consists of inventing a hypothetical greater-than-unity coupling constant (*12e*) that doesn't exist, and that when squared leads to a number greater than unity, which (so the argument goes) therefore should not count as a probability. The idea that the ability to imagine an abnormally large electromagnetic coupling constant should be counted as a refutation of a physical theory about our world leads to absurdities. One can also imagine, contrary to fact, that real photons have large finite rest mass, in which case photons would fail to travel on null cones. Does this mean that relativity is on shaky ground?

In any case, as is explicitly shown in Kastner and Cramer (2017), eqs (10) and (11), the square of the coupling amplitude between fields is only one factor in the total probability of a non-unitary transition. The relevant transition amplitudes contribute crucially to the probability that a measurement-type interaction will take place. These probabilities are decay rates, which depend on both the coupling constant and specific transitions between atomic states. Thus, transition probabilities are crucial aspects of the (time-dependent) probability of a measurement transition, and contribute factors that significantly decrease the basic coupling probability of 1/137.

The same observation applies in principle to the strong force coupling, in which the probability of non-unitarity would always be decreased by the relevant transition probabilities. Marchildon's suggestion that the strong coupling constant might exceed unity in no way refutes the interpretation of both Feynman and RTI concerning coupling amplitudes, since that situation only occurs for extreme separation between quarks, and expresses a critical transition zone, beyond the limit of quark confinement, in which enormous energies have to be injected. In this extreme zone, one has to put in so much energy that new quarks are created, which corresponds very nicely to exceeding what would be a coupling of unity for a single quark.

Thus, absorption and the advent of non-unitarity in RTI has indeed been quantitatively defined and remains so, whether or not one likes the interpretation of the square of the coupling amplitude (by itself) as a probability. Even if we were to reject that as a 'stand-alone' probability (and instead view it as only an important contribution), the total probability of the non-unitary measurement transition for a particular initial and final state is still as given in eqs. (10) and (11) of Kastner and Cramer (2017). That involves a standard, non-controversial result for a time-dependent probability, with the square of the coupling amplitude as an important contribution reflecting the basic tendency of charges to emit or absorb photons (which is what charge means). The relevant point is that emitters and absorbers

(e.g. atomic electrons) couple to the electromagnetic field, and that coupling is what corresponds to the possibility of OW and CW generation, all of this being duly quantified in the total transition probability for decays. The different factors of the decay probabilities have different physical significance: the square of the transition probability reflects the overlap between the initial and final states (given the electromagnetic perturbation), while the square of the coupling amplitude reflects the degree of potency or likelihood of basic OW/CW generation. It is not necessary that the latter strictly qualify as a 'stand-alone probability', though heuristically it appears to serve that purpose quite well. Certainly, all real-world coupling constants are less than or equal to unity (if the strong force exceeds unity, it reflects entry into the domain of quark creation as noted above), and the fact that one can dream up poorly behaved coupling constants that do not exist in the real world does not in any way demonstrate that the basic interpretation is flawed.

In conclusion, RTI has indeed provided quantitative specification of the required conditions for absorption and non-unitarity (the square of the relevant coupling constant times the transition amplitude, on-shell photon(s), and satisfaction of conservation laws), clearly allowing it to define the measurement transition. Marchildon's characterization of the ability of RTI to define measurement as 'overstated' would seem to apply instead to his own objections.